# Chern insulators without band inversion in MoS$_2$ monolayers with 3d adatoms

Xinyuan Wei[1], Bao Zhao[1,2], Jiayong Zhang[1], Yang Xue[1], Yun Li[3], and Zhongqin Yang[1,4]*

[1]*State Key Laboratory of Surface Physics and Key Laboratory for Computational Physical Sciences (MOE) & Department of Physics, Fudan University, Shanghai 200433, China*

[2] *College of Physical Science and Information Technology, Liaocheng University, Liaocheng 252000, China*

[3] *Department of Physics and Electronic Engineering, Hanshan Normal University, Chaozhou 521041, China*

[4]*Collaborative Innovation Center of Advanced Microstructures, Fudan University, Shanghai 200433, China*

Electronic and topological properties of MoS$_2$ monolayers endowed with 3d transition metal (TM) adatoms (V-Fe) are explored by using *ab initio* methods and *k p* models. Without the consideration of the Hubbard U interaction, the V, Cr, and Fe adatoms tend to locate on the top of the Mo atoms, while the most stable site for the Mn atom is at the hollow position of the Mo-S hexagon. After the Hubbard U is applied, the most stable sites of all the systems become the top of the Mo atoms. Chern insulators without band inversion are achieved in these systems. The V and Fe adsorption systems are the best candidates to produce the topological states. The *k p* model calculations indicate that these topological states are determined by the TM magnetism, the C$_{3v}$ crystal field from the MoS$_2$ substrate, and the TM atomic spin-orbit coupling (SOC). The special two-meron pseudospin texture is found to contribute to the topology. The apparent difference between the Berry curvatures for the V and Fe adsorption systems are also explored. Our results widen the understanding to the Chern insulators and are helpful for the applications of the MoS$_2$ monolayers in the future electronics and spintronics.

**Corresponding Author**

*E-mail: zyang@fudan.edu.cn

# I. INTRODUCTION

Transition metal dichalcogenide (TMD) thin films, graphene-like two-dimensional (2D) materials, have received immense attention recently due to their remarkable physical and chemical properties [1-2]. Due to the weak van der Waals interlayer bond, the TMD monolayers (MLs) can be easily exfoliated from the bulk materials. Through the exfoliating, the electronic structures can be varied dramatically. For example, the $MoS_2$ ML has a direct band gap of ~1.8 eV [3], while the bulk $MoS_2$ is an indirect semiconductor with a band gap of 1.3 eV [4]. The direct band gap and the gap value located in the visible light range not only make the experimental optoelectronic measurements in the system straightforward but also enable the $MoS_2$ ML to be more suitable for the electronic and optoelectronic applications [5,6]. More importantly, the coexistence of the nontrivial topological states of the valley Hall effect and spin Hall effect were predicted in the pristine $MoS_2$ ML due to the inversion symmetry breaking and spin-orbit coupling (SOC) in the system [7,8]. Therefore, the $MoS_2$ ML is expected to be an ideal material for valleytronics [9]. Based on the advantage of the optoelectronic measurements, valley polarization in the $MoS_2$ ML has been observed in experiments through the optical technologies, such as valley-selective circular dichroism [10,11].

Despite the spin Hall effect and valley Hall effect found in the $MoS_2$ ML, the possible realization of Chern insulators, namely quantum anomalous Hall (QAH) insulators [12], in $MoS_2$ MLs and other layered TMD materials is also a very interesting problem to be explored. Plenty of theoretical proposals have been raised to produce the QAH effect, including magnetic atom doped topological insulator films [13,14], graphene (silicene) based systems [15-19], and 2D organic material systems [20]. Fantastically, the experimental realization of the QAH effect was reported in the magnetic thin films of Cr- or V-doped $(Bi,Sb)_2Te_3$ topological insulators [21,22]. The forming mechanism of the QAH gap was primarily ascribed to the band inversion, in which the inversion of the valence and conduction bands with the same spin states (up-up or down-down) [13] or different spin states (up-down or down-up) [14] occurs. A number of the Chern insulators based on the graphene, silicene, and some other materials [15-19,23] can be understood by the band inversion mechanism proposed in Ref. [14], where the band inversion is primarily caused by the magnetic exchange field and the SOC merely opens the gap of the system. It is of great significance to investigate the Chern insulators beyond the band inversion mechanism.

In this work, we explore the electronic structures and the topological states of the $MoS_2$ ML with 3d transition metal (TM, TM=V, Cr, Mn, and Fe) atoms adsorbed. Chern insulators with parabolic band dispersions are found in these systems. The forming mechanism of the Chern insulators is different from the traditional band inversion. Without the SOC, the TM 3d orbitals exhibit degeneracy ($d_{x^2-y^2}$ and $d_{xy}$; $d_{xz}$ and $d_{yz}$), determined by the $C_{3v}$ symmetry from the crystal field of the $MoS_2$ substrate. Large nontrivial topological gaps are opened at the degenerate points by the TM atomic SOC. V and Fe adsorbed systems are found to be the suitable candidates to obtain the Chern insulators. The obviously asymmetrical behavior of the Berry curvature distribution for the Fe adsorbed case is ascribed to the large Rashba SOC interaction in the system.

## II. METHODS AND MODELS

Our calculations are performed by using density-functional theory (DFT) implemented in the Vienna *ab initio* Simulation Package (VASP) code [24]. The exchange-correlation interaction is described with the projector-augmented-wave (PAW) [25,26] method and the Perdew-Burke-Ernzerhof (PBE) approximation. Semicore pseudopotentials for Mo ($4p^6 4d^5 5s^1$), V ($3p^6 3d^3 4s^2$), Cr ($3p^6 3d^5 4s^1$), Mn ($3p^6 3d^5 4s^2$), and Fe ($3p^6 3d^6 4s^2$) are used to get accurate description of atomic potentials. The kinetic energy cutoff of plane waves is set as 400 eV. $8 \times 8 \times 1$ Monkhorst-Pack k-point meshes are adopted during the calculations. A $3 \times 3$ supercell is adopted (Fig. 1(a)), with the in-plane lattice constant fixed to the experimental value (3.16 Å) of the $MoS_2$ ML [27]. The calculations for the different supercell sizes are also performed. The $MoS_2$ MLs in adjacent supercell are separated by a vacuum region of at least 18 Å along the z direction to eliminate the undesired interactions. The total energy convergence criterion is set to be $10^{-5}$ eV per supercell. All atoms are fully relaxed until the Hellmann-Feynman forces are smaller than 0.01 eV/Å. Because the correlation effect is important in determining the electronic and magnetic properties of 3d TM atoms, the GGA+U method is considered with parameters of U=3 eV and J=0.9 eV in the calculations [28]. Three typical high-symmetry adsorption sites are chosen, as shown in Fig. 1(a), which are H (Hollow), M (Mo-top), and S (S-top). Other sites, such as the bridge site between the Mo and S atoms, are found not stable, on which the TM atom will transfer to any of the three high-symmetry sites.

## III. RESULTS AND DISCUSSION

### A. Adsorption analysis

The stability of 3d TM atoms adsorbed on the $MoS_2$ ML is studied at first. The binding energy is defined as: $E_b = E_{MoS_2} + E_{adatom} - E_{total}$, where $E_{MoS_2}$, $E_{adatom}$, and $E_{total}$ are the total energies of the pristine $MoS_2$ ML, the single TM atom, and the $MoS_2$ ML with the TM atoms adsorbed, respectively. In the calculations, the $E_{adatom}$ is taken as the total energy of the corresponding isolated TM atom. As shown in Table I, for each adsorption system (TM=V, Cr, Mn, or Fe), the binding energy values for the H and M sites are close to each other, which are much larger than that of the S site. Namely, the S adsorption site is the most unstable site. This tendency can be ascribed to the relative strong bonds formed not only between TM and S atoms but also between TM and Mo atoms in the first two cases, while only strong bonds between TM and S atoms can be formed in the S adsorption sites, confirmed by the very large values of the $d_{TM-Mo}$ (>4.2 Å) in the S adsorption sites. Between the H and M adsorption sites, the V, Cr, and Fe adatoms all tend to reside on the top of the Mo atoms (M sites) rather than the H sites, nevertheless the Mn adatom likes the H site, with the very small $E_b$ difference between the H and M adsorption sites. The results of these most stable adsorption sites are shown in bold in Table I. These obtained most stable adsorption sites are consistent with the results of Ref. [29], despite the different adsorption concentration considered. For these most stable adsorption sites (Table I), the Cr and Mn systems have smaller binding energies, meaning that the bonds between the adatom (Cr or Mn) and substrate are not as strong as the V or Fe cases. This can be understood by the more stable half-filling valence electron configurations of Cr and Mn atoms before the adsorption, given in the parenthesis in the last column of Table I. The total magnetic moment of each adsorption system with the most stable configuration can be rationalized by the valence electron configuration of the TM atom after the adsorption (Table I). Note that the interactions from the $MoS_2$ ML make the electrons transfer from 4s into 3d states for the V, Mn, and Fe systems, as also reported in graphene with some TM adatoms [30]. Thus, the total magnetic moment of V is very large (5.0 μB), while that of Fe is very small (2.0 μB).

When the Hubbard U term is applied, the most stable sites for the all four systems are the Mo-top, also given in Table I (the results of other adsorption sites with the U interaction are not shown). The U interaction makes the TM atoms more localized, resulting in the decrease of the binding energies of the all four systems. Reasonably, the adsorption heights and bond lengths are all elongated by the U interactions. It is interesting to find that the total magnetic moment of the Mn system increases from 3.0 μB to 5.0 μB after the U is considered due to the recovery of the valence electron configuration of the isolated Mn atom. Actually, the electronic structures of the $MoS_2$ ML with TM atoms doped were studied by several other groups [29,31,32]. Various theoretical models were adopted in these calculations [29,31,32], such as with different 3d or 4d TM atoms and different doping concentrations, etc. Some divergences can be found from these literatures with the same theoretical models. For example, the most stable adsorption sites and the total magnetic moments for V and Cr systems obtained in Ref. [31] are inconsistent with those given in Refs. [29,32]. These divergences may be associated with the different initial heights set for the TM atoms in the structural optimization calculations since our calculations show that the final states strongly depend on the initial height set. Thus, a series of different initial adsorption heights need to be set and calculated. More importantly, the Hubbard U interaction is not considered in these literatures, which is very important to produce the correct description of the electronic structures of systems containing TM atoms [28]. Its influences on the densities of states (DOSs) and the energy bands are sometimes very drastic, to be seen in the following.

The 3d and 4s partial DOSs of V, Cr, Mn, and Fe atoms adsorbed on the most stable sites of the $MoS_2$ ML are shown in Fig. 2. The left and right columns of Fig. 2 give the results without and with the U term considered, respectively. The light gray curve in each panel displays the total DOS of the supercell. Obviously, the occupied 3d states of the V and Cr atoms are mainly located within the band gap of the $MoS_2$ substrate, while the occupied 3d states of the Mn and Fe atoms are more extended, which overlap much with the $MoS_2$ bands, especially with the consideration of the U interaction. The symmetry of the system lowers from $D_{6h}$ of the bulk case to $C_{3v}$. Thus, the TM 3d orbitals split into three groups: one singlet state of $A$ ($d_z^2$) and two double degenerate states of $E_1$ ($d_{xz}$, $d_{yz}$) and $E_2$ ($d_{xy}$, $d_{x^2-y^2}$), marked in Fig. 2. For the V case (Fig. 2(a) and (b)), the spin-up states of $E_2$, $E_1$, and A are primarily located below the Fermi level ($E_F$), while the spin-down states are all above the $E_F$, leading to the half-filling of the V 3d orbitals. This behavior together with the unoccupied 4s orbitals (the green curves) gives the valence electron configuration of the V atom in this system of $3d^5 4s^0$, which confirms the electron transferring from the V 4s to 3d orbitals. These results are in good agreement with above analysis. The partial DOSs in other three systems can be understood similarly. The TM valence electron configurations in Fig. 2 are all consistent well with those listed in Table I. It is very significant to find that the Hubbard U interaction generally varies the TM partial DOSs drastically. Especially, for the Mn case, when the Hubbard U is turned on, the spin-up 3d states are pushed downwards in energy, while the spin-down 3d states are moved upwards, resulting in the completely unoccupied spin-down 3d states. The two electrons in the original spin-down 3d states transfer to the Mn 4s states under the U interaction, as seen by comparing the Mn 4s DOS distributions in Fig. 2(e) and (f). The essential reasons for these variations are the stability of the 3d half-filled orbitals and the increased localization of the Mn 3d electrons after the U interaction is turned on.

## B. Band structures and Berry curvatures

The band structures of the MoS$_2$ ML with V, Cr, Mn, or Fe adatoms are shown in Fig. 3, where the Hubbard U interaction is applied. The upper and lower rows correspond to the bands without and with the considerations of the SOC, respectively. We first discuss the bands without the SOC. The dominating components of the TM bands within the substrate band gaps have been marked in Fig. 3(a)-(d). One common behavior of these marked bands is the parabolic band dispersion near the Γ point. The two-fold degeneracy of E$_1$ and E$_2$ bands in V, Cr, and Fe cases can also be clearly seen, determined by the C$_{3v}$ crystal field from the substrate. Note that the spin-up E$_1$ and E$_2$ bands for the Mn (Fe) cases are not located within the substrate gap, but deeply below the E$_F$, as seen from Fig. 2(f) (2(h)). Since the Fe 3d states in Fig. 3(d) are over half-filled (3d$^8$4s$^0$), the spin-down A and E$_2$ bands are located near the E$_F$. When the SOC is taken into account in these systems, apparent gaps are opened at the Γ points, the original degenerate points, of the E$_2$ bands in the V, Cr, or Fe systems. The degeneracy at the Γ points of the E$_1$ bands in the V or Cr systems is actually also lifted by the TM atomic SOC despite the very small gaps opened.

We now take the E$_1$ and E$_2$ bands of the V system (Fig. 3(e)) and the E$_2$ band of the Fe system (Fig. 3(h)) as examples to explore their possible nontrivial topological properties. The corresponding enlarged figures are displayed in Fig. 4. The band gaps opened at the Γ points by the atomic SOC in Fig. 4 (a)-(c) are 22.4, 8.6, and 67.9 meV, respectively, much larger than the usual gaps opened with band inversion mechanism [15-19]. And global gaps exist in Fig. 4(a) and (c), while there is no global gap for the E$_1$ band in Fig. 4(b). Hence, semimetal bands are found in Fig. 4(b) if the E$_F$ is tuned to about -0.26 eV. Note that in the Fe case, the minimum gap of the system is not at the Γ point, but at about the midpoint in the k path between the K and Γ points.

To identify the topological properties of these SOC induced band gaps, Berry curvatures and Chern numbers are calculated by using maximally localized Wannier functions (MLWF) [33-35]. The Berry curvature results are shown by blue dashed curves in Fig. 4(a)-(c). Due to the absence of the inversion symmetry in the systems, the signs of the Berry curvatures at the K and K' points are opposite [8,9] for all the three cases. Thus, their contributions to the Chern number of the system are cancelled to certain extent. The distributions of the Berry curvatures around the $\Gamma$ points of the three systems are substantial, which contribute much to the net Chern numbers. Anyhow the Chern numbers obtained for Fig. 4(a)-(c) are -1, -1, and 1, respectively, which give a direct evidence of the Chern insulating gaps opened in these systems. The edge states of the $MoS_2$ nanoribbons with V or Fe adatoms were calculated by using Wannier function constructed tight binding model. As shown in Fig. 4(d-f), the metallic edge states (blue curves) do exist for each case. The two chiral edge states of each case correspond to the states located along the two edges of the nanoribbons with finite widths, respectively. Namely, one chiral conducting channel will appear on each side of the sample. The trend is consistent with the results of the Chern number ($|C|=1$) obtained in these cases and further confirms the existence of the nontrivial topological properties in the systems.

In Fig. 4(c), it is interesting to find that around the $\Gamma$ point, only one peak of the Berry curvature exists, located at the position of the minimum band gap of the system, while there are two peaks (with minus values) in Fig. 4(a) and (b), which will be discussed in the following. The forming reasons of these Chern insulating gaps are very different from the band inversion mechanism mentioned above [15-19,23] despite some also with the parabolic band dispersions at the $\Gamma$ point around the $E_F$ [13,14] due to no band inversion happening in the gap opening process in these systems. These $E_1$ or $E_2$ bands are originally degenerate at the $\Gamma$ points. The degeneracy is lifted by the TM atomic SOC and meantime the nontrivial band gaps are opened. Therefore, the forming mechanism of these Chern insulators is quite novel. In some QAH systems reported previously [20,36], the band inversion was also not found. The dispersion of the concerned bands in them is the traditional linear Dirac type instead of the parabolic bands found in our systems. Besides, the Chern insulating gaps achieved here are formed directly by the 3d states of the TM adatoms, tuned by the crystal field from the $MoS_2$ ML substrate, which is also different from many previous reports of Chern insulating gaps formed primarily by the states of the substrates [15,18] or the hybrid states of the substrates and the adatoms [37]. The long range ferromagnetic order in these $MoS_2$ monolayers with 3d adatoms, essential for the QAH state, may be established by the so called Van Vleck paramagnetism mechanism [14], provided with an estimated sizable Van Vleck magnetic susceptibility of $MoS_2$ materials [38]. This mechanism is very different from that of the traditional dilute magnetic semiconductors due to no enough free carriers to mediate the magnetic interaction in the system. Based on the method introduced in Ref. [14], the Curie temperature is roughly estimated to be about 35 K for the $MoS_2$ with Fe adsorbed [39].

Thus, we could draw a conclusion here that the Chern insulators can be achieved well in the V or Fe adsorbed $MoS_2$ ML if the $E_F$ is adjusted to the $E_2$ gaps of the systems. Due to no global gap opened for the $E_1$ state in the V system, only a Chern semimetal can be obtained in this case. The 2D Berry curvature distributions in the momentum space corresponding to the cases of Fig. 4(a) and (c) are given in Fig. 5. Very explicit $C_{3v}$ symmetry can be found in the 2D Berry curvature distributions of Fig. 5. The somewhat asymmetric distributions along K-Γ-K' of the Berry curvatures in Fig. 4 can be understood by this $C_{3v}$ symmetry. The electronic structures of the systems with various TM adsorption concentrations are also investigated by using different sizes of the supercell, such as 1×1, 2×2, and 4×4. It is found that the band gaps of the 1×1 systems usually cannot be opened, related to the strong direct interactions between the TM atoms. For supercells equal or larger than 2×2, the band gaps can be generally opened around the Γ points with the presence of SOC. The robust nontrivial topological properties are also confirmed with respect to the thickness of $MoS_2$ substrate by V adsorbed $MoS_2$ bilayer and trilayer calculations. The robustness of the obtained topological states facilitates the fabrication of the Chern insulators and the measurements of its properties in experiments. As shown in Fig. 4, the achieved nontrivial topological gaps are usually opened below the $E_F$. With proper hole doping, the bands can be shifted upwards and the $E_F$ can be located exactly within the nontrivial gap opened. For example, for $E_2$ bands in Fig. 4(c), the $E_F$ can be tuned to the $E_2$ gap by removing one electron from the system. And the hole doping concentration is about $1.1 \times 10^{14}$ cm$^{-2}$, which can be realized in current experiments with advanced gating technologies [40,41]. To observe the effect in experiments, the adatom distribution needs to be periodical on the MoS$_2$ surface and does not break the symmetry of the substrate ($C_{3v}$). Thus, high sample quality is required to try to lower the randomization effect of the adatoms in experiments. Although the binding energies for the V and Fe cases (with Hubbard U) are not small (~1 eV), the adatoms may still have certain probability to hop to other sites by overcoming the energy barrier (~0.5 eV) due to the thermal vibrations. Thus, to carry out easily this effect in experiments, low temperature condition needs to be satisfied, which is also required by the nontrivial gaps opened in the systems.

## C. *k p* model calculations

To better understand the topology origin in these systems, we derive an effective Hamiltonian in terms of the *k p* model for low-energy physics based on the invariant theory [42,43]. The $E_2$ bands, composing of $d_{x^2-y^2}$ and $d_{xy}$ orbitals, of the V and Fe adsorbed MoS$_2$ MLs are focused. For convenience, the two orbitals are transformed to $|d_{\pm 2}\rangle = |d_{x^2-y^2}\rangle \pm 2i|d_{xy}\rangle$ with definite orbital angular momentums.

In the absence of SOC and magnetism, the system contains time-reversal symmetry (T), mirror symmetry ($M_x$) along the *x* direction, and the threefold rotation ($C_3$) along the *z* axis. Under the basis of $|d_{+2},\uparrow\rangle$, $|d_{-2},\uparrow\rangle$, $|d_{+2},\downarrow\rangle$, and $|d_{-2},\downarrow\rangle$, the transformation matrices of the symmetry operations are $T = \kappa \cdot i\sigma_y \otimes \tau_x$, $M_x = -i\sigma_x \otimes \tau_x$, and $C_3 = \exp(-i\frac{2\pi}{3}J_z) \otimes 1$, where $\kappa$ represents complex conjugation operator, $\sigma_i$ and $\tau_i$ are the Pauli matrices in spin and band space, respectively, and $J_z$ is the angular momentum in the *z* direction. Thus, the *k p* model around the $\Gamma$ point can be constructed up to the second order of *k*:

$$H_0 = \alpha k^2 + \begin{pmatrix} 0 & \beta k_+^2 & 0 & 0 \\ \beta^* k_-^2 & 0 & 0 & 0 \\ 0 & 0 & 0 & \beta k_+^2 \\ 0 & 0 & \beta^* k_-^2 & 0 \end{pmatrix}, \tag{1}$$

with $k_\pm = k_x \pm i k_y$. For the atomic SOC effect, $H_{so} = \lambda_{so} L \cdot S$ is diagonal in the selected basis. The ferromagnetic exchange term is added as $H_M = M_{\sigma_z} \otimes 1$. So the total Hamiltonian can be written as:

$$H = H_0 + H_{so} + H_M \tag{2}$$

Diagonalizing the above Hamiltonian without the SOC and exchange field terms, we obtain the dispersion featured by spin degeneracy and quadratic band touching at the $\Gamma$ point, as displayed in Fig. 6(a). When the exchange field term is turned on, the spin-up and spin-down bands can be completely separated from each other, as shown in Fig. 6(b). When the SOC effect is further included, band gaps are opened at the quadratic degenerate points of the spin-up and spin-down bands as shown in Fig. 6(c) and (d), which correspond to the cases of Fig. 4(a) and (c), respectively. To check the topological properties of the SOC induced gap, we calculate the Berry curvatures along the high-symmetry directions and the Chern numbers when the $E_F$ is located at the SOC gap according to Kubo formalism [44,45] with this $k \cdot p$ model. As shown in Fig. 6(c) and (d), the Berry curvatures form peaks around the $\Gamma$ points (with minus values in Fig. 6(c)), which lead to the Chern numbers $C = -1$ and 1 for the spin-up and spin-down bands, respectively. These Chern numbers are also in agreement with those obtained in Fig. 4(a) and (c), respectively. Thus, the built $k \cdot p$ model explains well the topology obtained in the TM atom adsorbed MoS$_2$ ML. No band inversion is indeed needed in the process of the nontrivial gap opening.

To explore the topology in another point of view, we plot the pseudospin textures for the upper (spin-up bands) and lower (spin-down bands) blocks of the total Hamiltonian, as shown in Fig. 6(e), (f), respectively. There are obviously two-meron texture structures for each block. Since each meron texture will contribute a half topological charge [16,46], the two-meron texture should lead to one topological charge. By calculating the pseudospin Chern number as $n = 1/4\pi \int dk^2 (\partial k_x \hat{h} \times$ $\partial k_y \hat{h}) \cdot \hat{h}$, where $\hat{h} = h/|h|$ with $h$ representing the projection of the Hamiltonian in equation (2) into pseudospin space, we confirm that the pseudospin Chern number for the spin-up (down) bands is indeed -1(1). The pseudospin textures obtained here are different from those of the honeycomb lattice, where the two merons are located at the two inequivalent K and K' points, respectively [16,20]. The difference is associated with the quadratic band dispersion instead of the Dirac-cone-like bands obtained in our work.

**D. Rashba effect**

Dissimilar to Fig. 4(a) and (c), the Berry curvature distributions in Fig. 6(c) and (d) are symmetric around the $\Gamma$ points due to no lattice symmetry considered in the $k \cdot p$ model. We now explain why there is only one sharp peak of the Berry curvature in Fig. 4(c), also inconsistent with the result in Fig. 6(d). In general, the magnitude of the Berry curvature is inversely proportional to the nontrivial gap value [45]. Thus, the severely asymmetric band distribution around the $\Gamma$ point in Fig. 4(c) should be comprehended. Note that there is no such obvious phenomenon in the bands of Fig. 4(a). By comparing the $E_2$ bands in Fig. 3(d) and (h), one can see that the asymmetric behavior is triggered by the SOC interaction. As listed in Table I, the Fe atom has the lowest height ($h$) from the substrate, which can induce a strong Rashba SOC [47] in the system due to the strong asymmetric potential distribution on the two sides of the TM plane [41,48,49]. To confirm this speculation, the bands of the Fe adsorbed $MoS_2$ ML with the Fe height increasing to 1.5 Å are calculated and given in Fig. 7(a). Explicitly, the $E_2$ bands become more symmetric around the $\Gamma$ point, just like the V adsorption case (Fig. 4(a)). The charge distributions of the $E_2$ bands of the Fe cases with the normal (0.93 Å) and enlarged (1.5 Å) height are given in Fig. 8. When the Fe adatom is moved upwards and fixed at 1.5 Å above the upper S plane, the charge distributions of the Fe atom tend to be symmetric about its center and more look like that of an isolated Fe atom due to the less interaction from the substrate. The Rashba SOC is, thus, reduced much [48]. The band evolution under the atomic SOC and the Rashba SOC is illustrated in Fig. 7(b), where the influence of the Rashba SOC on the parabolic bands can be traced to analysis in Refs. [15,50]. Thus, the asymmetric behavior of the $E_2$ bands (corresponding to the spin-down bands in Fig. 7(b)) with respect to the $\Gamma$ point in the Fe case (Fig. 4(c)) can be understood very legibly. It is worth emphasizing that the topological properties keep undistributed with this Rashba effect, beneficial to the application of the $MoS_2$ system in spintronic devices. It is, of course, a very interesting topic to explore some other possible substrates with a triangular or honeycomb lattice to realize the QAH effect.

## IV. CONCLUSIONS

We systematically studied the electronic and topological properties of the $MoS_2$ monolayer with 3d TM (TM=V, Cr, Mn, and Fe) atoms adsorbed. The Hubbard U interaction is found having an important impact on the electronic and further the topological properties of the system. With the consideration of the Hubbard U, the most stable sites of all the systems studied become the Mo-top. Chern insulators can be well achieved in the V and Fe doped systems. The mechanism is differ from that in graphene and silicene etc., where the topology is mainly triggered by the band inversion mechanism. The Chern insulating gaps here are formed directly by the 3d states of the TM adatoms, tuned by the crystal field from the $MoS_2$ ML substrate. Without the SOC, the TM 3d orbitals exhibit parabolic dispersion and degeneracy at the $\Gamma$ point. Large nontrivial gaps can be opened in these degenerate bands by the TM atomic SOC. A strong Rashba SOC effect is found in the Fe adsorbed system, which causes very asymmetrical behaviors of the Berry curvature distribution. The reason of the large Rashba effect was analyzed.

## ACKNOWLEDGMENTS

The authors are grateful to R. Yu and X. Hu for very helpful discussion. This work was supported by National Natural Science Foundation of China under Grant Nos. 11574051 and 61171011, Natural Science Foundation of Shanghai with grant No. 14ZR1403400. X. W. acknowledged the support from China Scholarship Council (CSC). Y. L. thanked the support from Open Project of State Key Laboratory of Surface Physics at Fudan University. All the calculations were performed on the High Performance Computational Center (HPCC) of Department of Physics at Fudan University.

**Table I** The results for the MoS$_2$ ML with a 3d TM (TM=V, Cr, Mn, and Fe) atom adsorbed in a 3x3 supercell. The properties listed contain the adsorption site, binding energy ($E_b$), adsorption height relative to the upper S plane ($h$), bond length of the TM adatom to the nearest neighbor S ($d_{TM-S}$) and Mo ($d_{TM-Mo}$) atoms, total magnetic moment (Mag), and the valence electron configuration of the TM atom (Val. Conf.) in the adsorption system. The expression in the parenthesis in the last column indicates the valence electron configuration of the isolated TM atom. The rows beginning with 'M(+U)' indicate the results of the M adsorption site with the Hubbard U interaction applied. The rows in bold show the results of the most stable sites without or with the Hubbard U considered.

| TM | Site | $E_b$ (eV) | $h$ (Å) | $d_{TM-S}$ (Å) | $d_{TM-Mo}$ (Å) | Mag ($\mu_B$) | Val. Conf. |
|---|---|---|---|---|---|---|---|
| | H | 1.95 | 1.15 | 2.26 | 3.31 | 5.0 | |
| V | **M** | **2.46** | **1.25** | **2.29** | **2.87** | **5.0** | **3d$^5$4s$^0$(3d$^3$4s$^2$)** |
| | S | 1.22 | 2.28 | 2.28 | 4.27 | 5.0 | |
| | **M(+U)** | **1.02** | **1.30** | **2.32** | **2.90** | **5.0** | **3d$^5$4s$^0$** |
| | H | 0.99 | 1.61 | 2.46 | 3.69 | 6.0 | |
| Cr | **M** | **1.16** | **1.51** | **2.41** | **3.11** | **6.0** | **3d$^5$4s$^1$(3d$^5$4s$^1$)** |
| | S | 0.59 | 2.43 | 2.43 | 4.41 | 6.0 | |
| | **M(+U)** | **0.30** | **1.69** | **2.51** | **3.28** | **6.0** | **3d$^5$4s$^1$** |
| | **H** | **1.03** | **0.85** | **2.15** | **3.09** | **3.0** | **3d$^7$4s$^0$(3d$^5$4s$^2$)** |
| Mn | M | 0.99 | 1.01 | 2.20 | 2.64 | 3.0 | |
| | S | 0.27 | 2.36 | 2.36 | 4.35 | 5.0 | |
| | **M(+U)** | **0.14** | **1.81** | **2.59** | **3.40** | **5.0** | **3d$^5$4s$^2$** |
| | H | 2.27 | 0.67 | 2.09 | 2.94 | 2.0 | |
| Fe | **M** | **2.43** | **0.90** | **2.14** | **2.55** | **2.0** | **3d$^8$4s$^0$(3d$^6$4s$^2$)** |
| | S | 0.96 | 2.22 | 2.22 | 4.22 | 4.0 | |
| | **M(+U)** | **0.87** | **0.93** | **2.16** | **2.57** | **2.0** | **3d$^8$4s$^0$** |

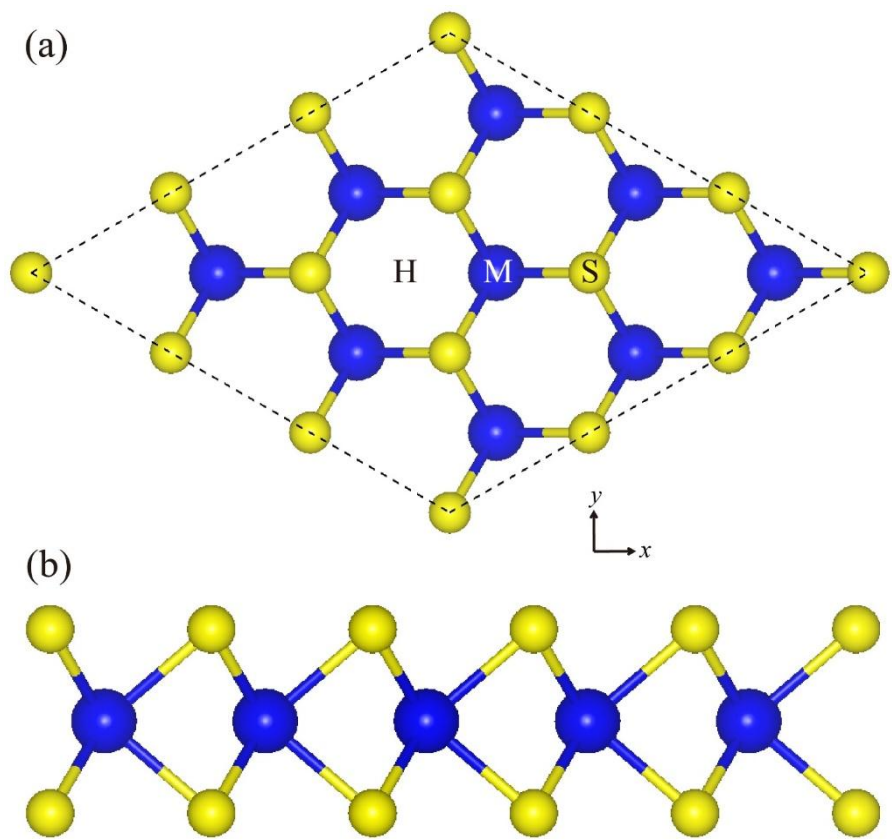

**Figure 1** (Color online) Top (a) and side (b) views of the $MoS_2$ ML. Three high-symmetry adsorption sites, labeled as H (Hollow), M (Mo-top), and S (S-top), are shown in (a). The blue and yellow spheres represent the Mo and S atoms, respectively. The dashed lines illustrate the $3\times3$ supercell.

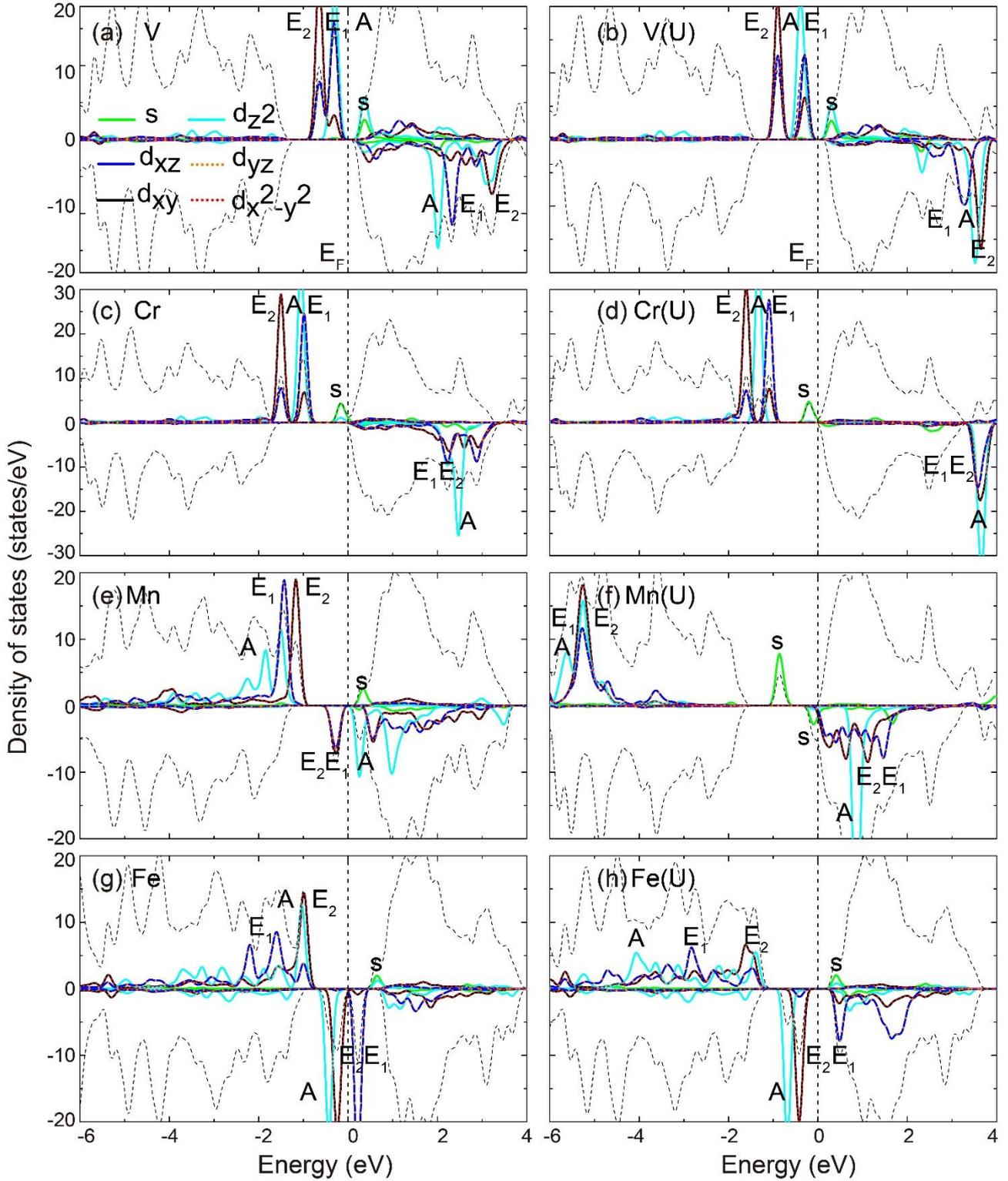

**Figure 2** (Color online) The 3d and 4s partial DOSs of the TM atoms adsorbed on the MoS$_2$ ML. For clarity, the 3d and 4s partial DOSs are amplified by a factor of 10. The light gray curves give the total DOSs of the supercells. The peaks which marked with E$_1$, E$_2$, and A are the 3d orbital components for the TM adatoms under the C$_{3v}$ crystal field. The positive and negative DOS values

correspond to the spin-up and spin-down states, respectively. The vertical dashed line in each panel indicates the Fermi level. The left and right columns correspond to the results without and with the Hubbard U considered, respectively.

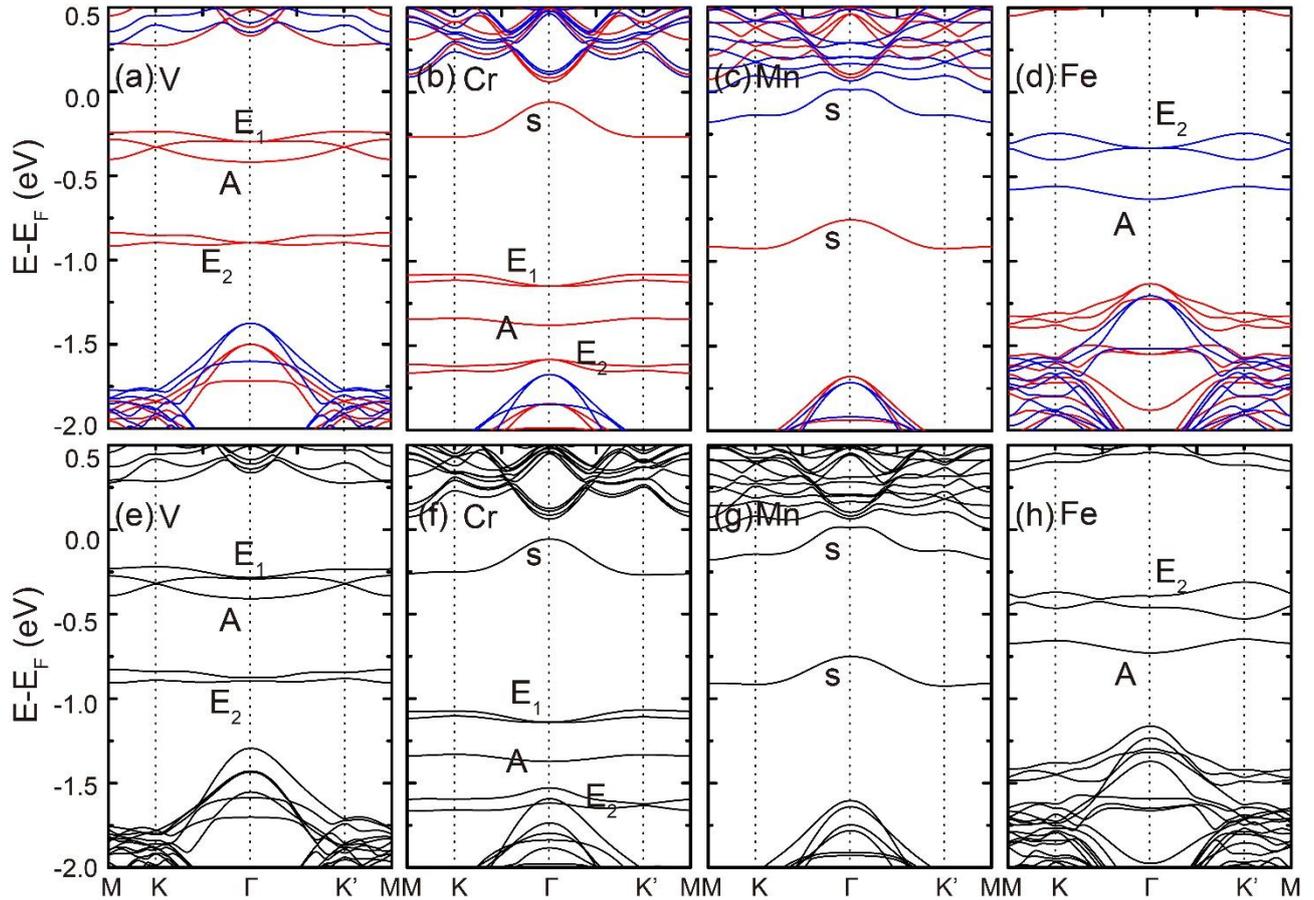

**Figure 3** (Color online) The band structures of the $MoS_2$ ML with V, Cr, Mn, or Fe adsorbed. The Hubbard U interaction is considered. The red and blue curves denote the spin-up and spin-down components, respectively. The upper and lower rows give the results without and with the consideration of the SOC. The $E_F$ is set as zero.

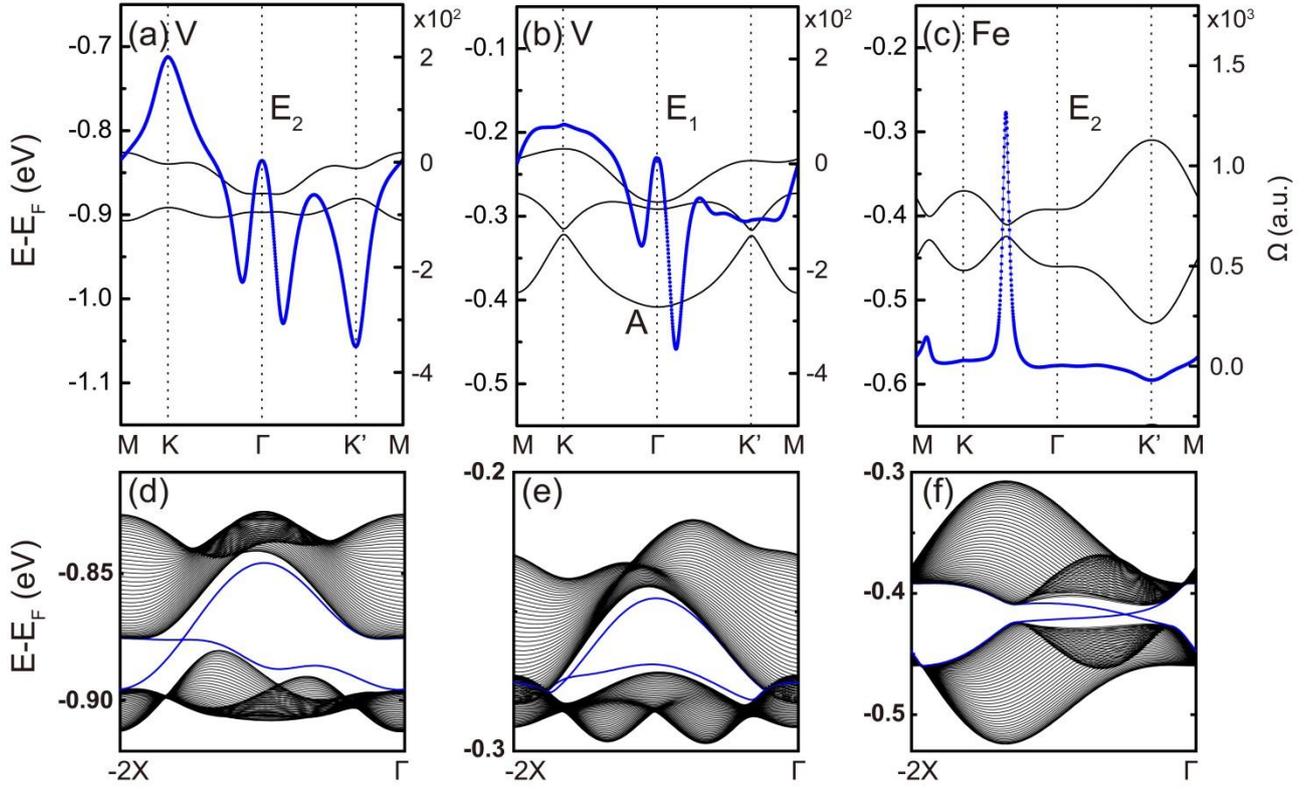

**Figure 4** (Color online) (a) and (b) the enlarged $E_2$ and $E_1$ bands in Fig. 3(e), respectively. (c) the enlarged $E_2$ band in Fig. 3(h). The blue dashed curves denote the corresponding Berry curvatures ($\Omega$) in atomic units (a.u.). The corresponding surface states of the nanoribbions with finite widths are given in (d), (e), and (f), respectively.

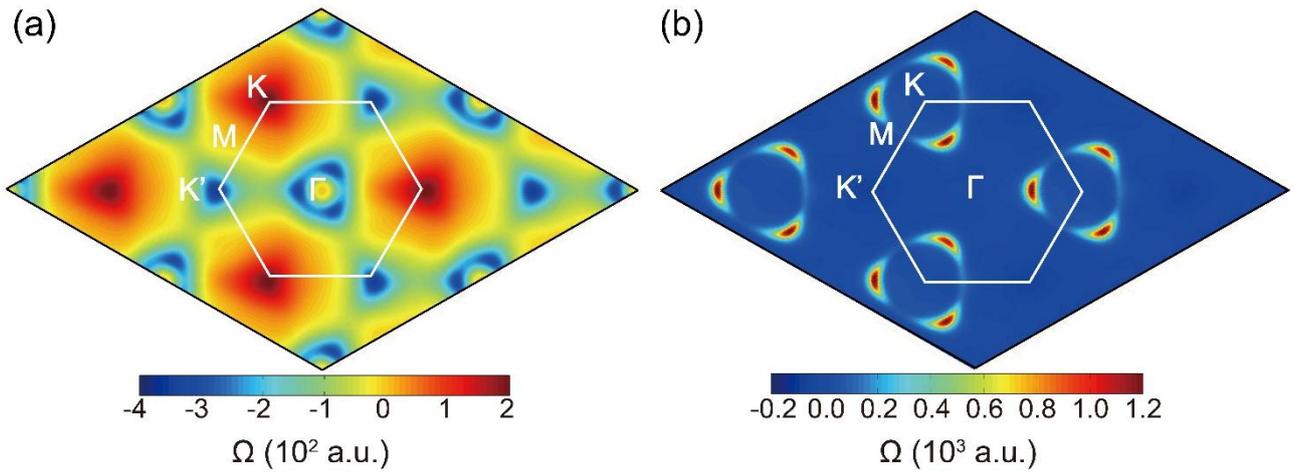

**Figure 5** (Color online) The 2D Berry curvature distributions (in a.u.) in the momentum space of the $E_2$ bands in the V (a) and Fe (b) adsorbed $MoS_2$ MLs. The four high-symmetry k points are given.

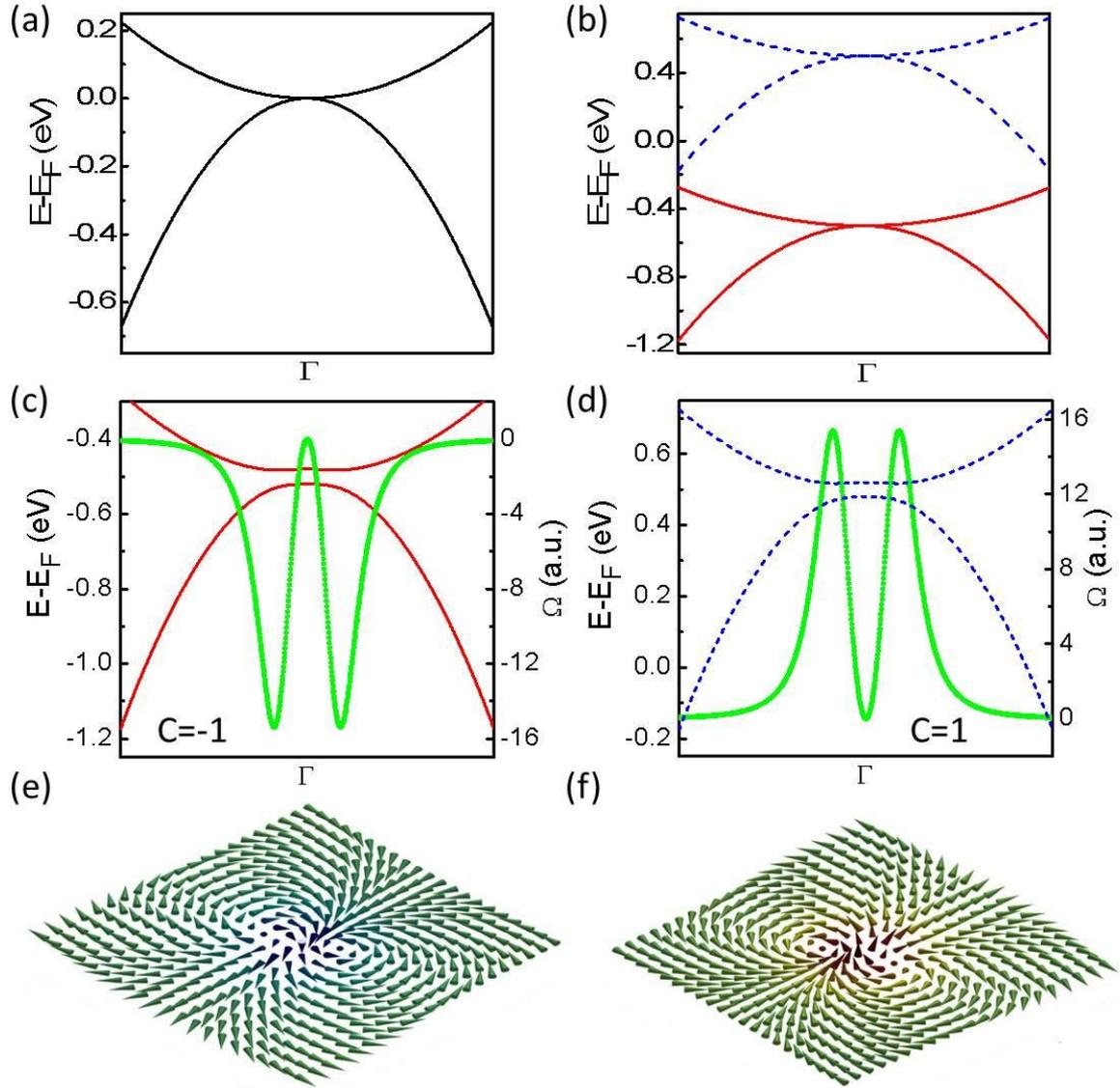

**Figure 6** (Color online) The *k p* model band structures around the Γ point under the parameters (a) α= -0.1 eV·Å², β= 0.2 eV·Å, M = 0.0 eV, λso= 0.0 eV; (b) α= -0.1 eV·Å², β= 0.2 eV·Å, M = -0.5 eV, λso= 0.0 eV; (c) and (d) α= -0.1 eV·Å², β= 0.2 eV·Å, M = -0.5 eV, λso= -0.02 eV for the spin-up and spin-down bands, respectively. The red solid curves (blue dashed curves) represent the spin-up (spin-down) bands. With the $E_F$ located within the SOC induced gap, the obtained Berry curvatures (green solid curves) and the Chern numbers C for the spin-up and spin-down bands are also given in (c) and (d). (e) and (f) are the pseudospin textures around the Γ points for the spin-up and spin-down bands, respectively.

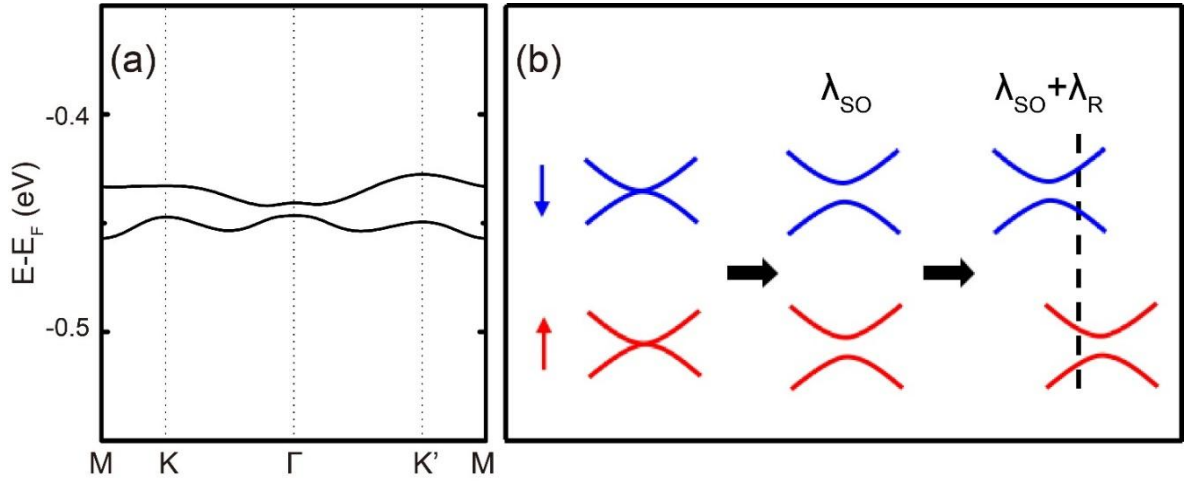

**Figure 7** (Color online) (a) The corresponding bands of Fig. 4(c) with the height (*h*) of Fe adatom increasing from 0.93 Å to 1.5 Å. (b) Schematic diagram of the band evolution with the consideration of the atomic SOC ($\lambda_{so}$) and Rashba SOC ($\lambda_R$).

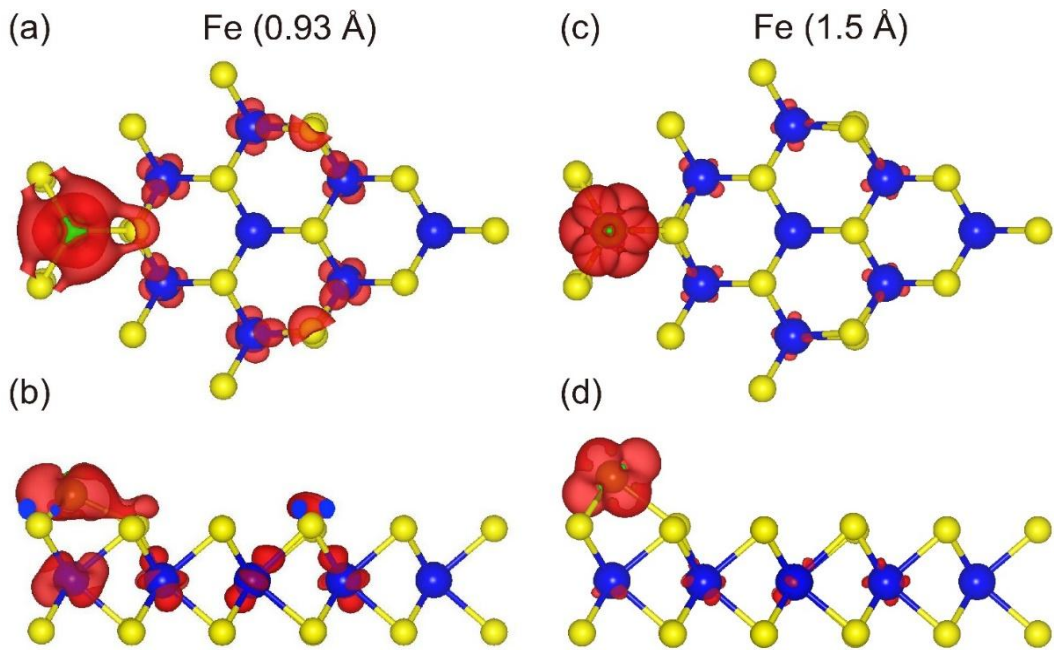

**Figure 8** (Color online) The charge distributions of the $E_2$ bands in the Fe cases with the Fe atom at the normal height (a,b) or enlarged height (c,d). The upper and lower panels give the top and side views, respectively. The isosurface value is 0.002 $e/a_0^3$.